\documentclass{pr-imfp06}

\def \dzero    {D\O}
\def \pt       {p_{\mathrm{T}}}

\def \Et       {E_{\mathrm{T}}}
\def \met   {\not\!\!\Et}

\def \dzero    {D\O}

\def \ttbar    {t\bar{t}}
\def \ppbar    {p\bar{p}}
\def \qqbar    {q\bar{q}}
\def \invpb    {\mathrm{\,pb^{-1}}}

\begin{document}

\title{Looking for single top quarks in \dzero~data}

\author{Aran Garcia-Bellido}

\address{Physics Department, University of Washington, BOX 351560 \\ 
Seattle, WA 98195-1560, U.S.A.\\
E-mail: aran@fnal.gov}

\maketitle

\abstracts{The search for the electroweak production of top quarks is entering
an exciting time: with the ever increasing luminosity furnished by the Tevatron 
$\ppbar$ Collider at Fermilab and sophisticated analysis techniques we are now
able to approach the expected standard model production cross section for this
new mode of production. Using 370$\invpb$ of \dzero~data and likelihoods as
discriminants to extract the signal from the large backgrounds, we set upper
limits on the standard model production cross section in the $s$-channel of
5.0~pb and in the $t$-channel of 4.4~pb, at 95\% confidence level.}

\section{Introduction}
\label{sec:intro}
\begin{figure}
\begin{center}
  \includegraphics[width=0.35\textwidth]{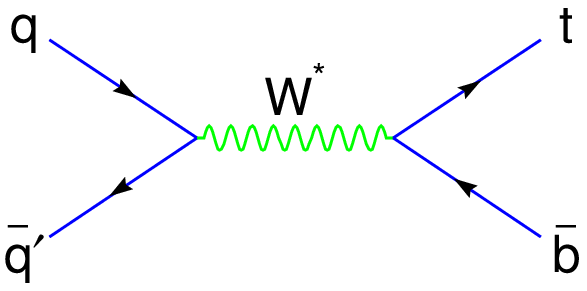}
  \hspace{0.2cm}
  \includegraphics[width=0.35\textwidth]{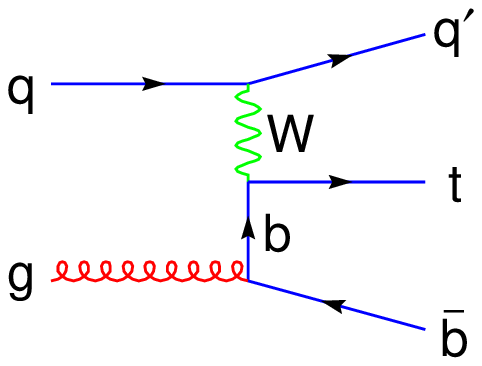}
  \caption{The two dominant production modes of single top quarks at the Tevatron:
$s$-channel (left) and $t$-channel (right) exchange of a $W$ boson.}
\label{fig:feyn}
\end{center}
\end{figure}

Top quarks have been observed so far produced in $\ttbar$ pairs in collisions at
Fermilab's proton-antiproton Tevatron Collider. Pair production occurs
predominantly when two incoming quarks exchange a gluon ($\qqbar\to g \to\ttbar$):
that is, they are 
mediated by the strong force. But another mode of production, which has not been
detected yet, is predicted by the standard model (SM), when the incoming quarks
exchange a $W$-boson and as a result a single top quark is produced along with a
$b$-quark or other light quarks. This is an electroweak top quark production,
also referred as single top quark production.
There are two main modes of
$W$~boson exchange with top quark production observable at the Tevatron:
$s$-channel and $t$-channel, as depicted in Fig.~\ref{fig:feyn}. At NNLO, the
predicted cross sections are 0.88 and 1.98~pb respectively. 

The observation and study of this new production mechanism is important for the
following reasons: a measurement of the cross section can yield the first direct
measurement of the CKM matrix element $|V_{tb}|$ without any assumptions on the
number of families in the SM; new particles can enhance the production cross
section in the $s$-channel, and anomalous couplings can alter the production
through the $t$-channel, thus this channel offers a very sensitive window into
new physics; this channel can be used to test the V-A nature of the SM; and not
least important, this channel requires a good knowledge of the W+jets background
and is itself a background to Higgs boson searches.  For more details of the analyses
described here, see Ref.~\cite{d0singletop}. 

\section{Signal and Background Topologies}
The signal topology we are interested in arises from the top quark decaying to
$Wb$ and the $W$ decaying leptonically to electrons or muons. We are
looking for one isolated high $\pt$
lepton, missing transverse energy, and at least one $b$-quark jet (two in the
case of the $s$-channel, and usually only one in the $t$-channel, since the $b$
from gluon splitting tends to get lost in the beam-pipe). The main background is W+jets
production: these events tend to be a little bit more pencil-like and tend to
have smaller total reconstructed energy. But the cross section for W+jets is at
least two orders of magnitude bigger than that of single top. The second largest
background is $\ttbar$ production in the $\ell$+jets or dilepton decay
modes. These events tend to have more total reconstructed energy (having two
heavy top quarks in them) and be more spherical than single top. The $\ttbar$
cross section is around twice that of single top. 

The $W$+jets background is modeled using Alpgen~\cite{alpgen}, but it is
normalized to the data. The $\ttbar$ background is also modeled using Alpgen and
normalized to the NNLO cross section of 6.7~pb. The QCD background
is obtained from data and the fraction present in the signal data is obtained
from the same normalization of W+jets to signal data. For single top, we employ
CompHEP~\cite{comphep}, which effectively describes all angular correlations
between the final state particles at NLO. 

\section{Analyzed Data and Event Selection}
Data collected with the \dzero~detector~\cite{d0nim} between August 2002 and October 2004
corresponding to nearly 370$\invpb$ have been analyzed in the search for $s$-
and $t$-channel single top quark signals.

We apply loose selection cuts to the data and the Monte Carlo to extract
signal-like physics: we require to have 2, 3 or 4 jets in the event, the leading
one with a $p_{T}>25$~GeV and $|\eta|<2.5$, the rest with $p_{T}>15$~GeV and
$|\eta|<3.4$; we additionally require one isolated electron or muon with $\pt >
15$~GeV and $|\eta(e)|<1.1$ or $|\eta(\mu|)<2.0$; and the transverse missing energy
to be $\met>15$~GeV. 

Finally, we require the presence of at least one jet identified as coming from
a $b$ quark. We use the Jet LIfetime Probability (JLIP) algorithm based on the fact that
jets originated from a $b$ quark exhibit tracks with large impact parameters in
the transverse plane and thus allow to separate jets originated close to the
primary vertex or not. For each jet we have a continuous distribution and decide
to ``tag'' the jet as a \emph{tight} $b$ jet if $P_{JLIP}<0.3\%$ or \emph{loose}
$b$ jet if $P_{JLIP}<1\%$. The samples are separated into events with one tight
and no loose $b$ jets, and events with one tight $b$ jet and one loose
$b$ jet. Allowing the second $b$ tagged jet to be looser than the first,
increases our acceptance and somewhat reduces the uncertainty coming from $b$
tagging in these small samples. 

\section{Discriminant Variables and Likelihood discriminants}
\begin{figure}
\begin{center}
  \includegraphics[width=0.36\textwidth]{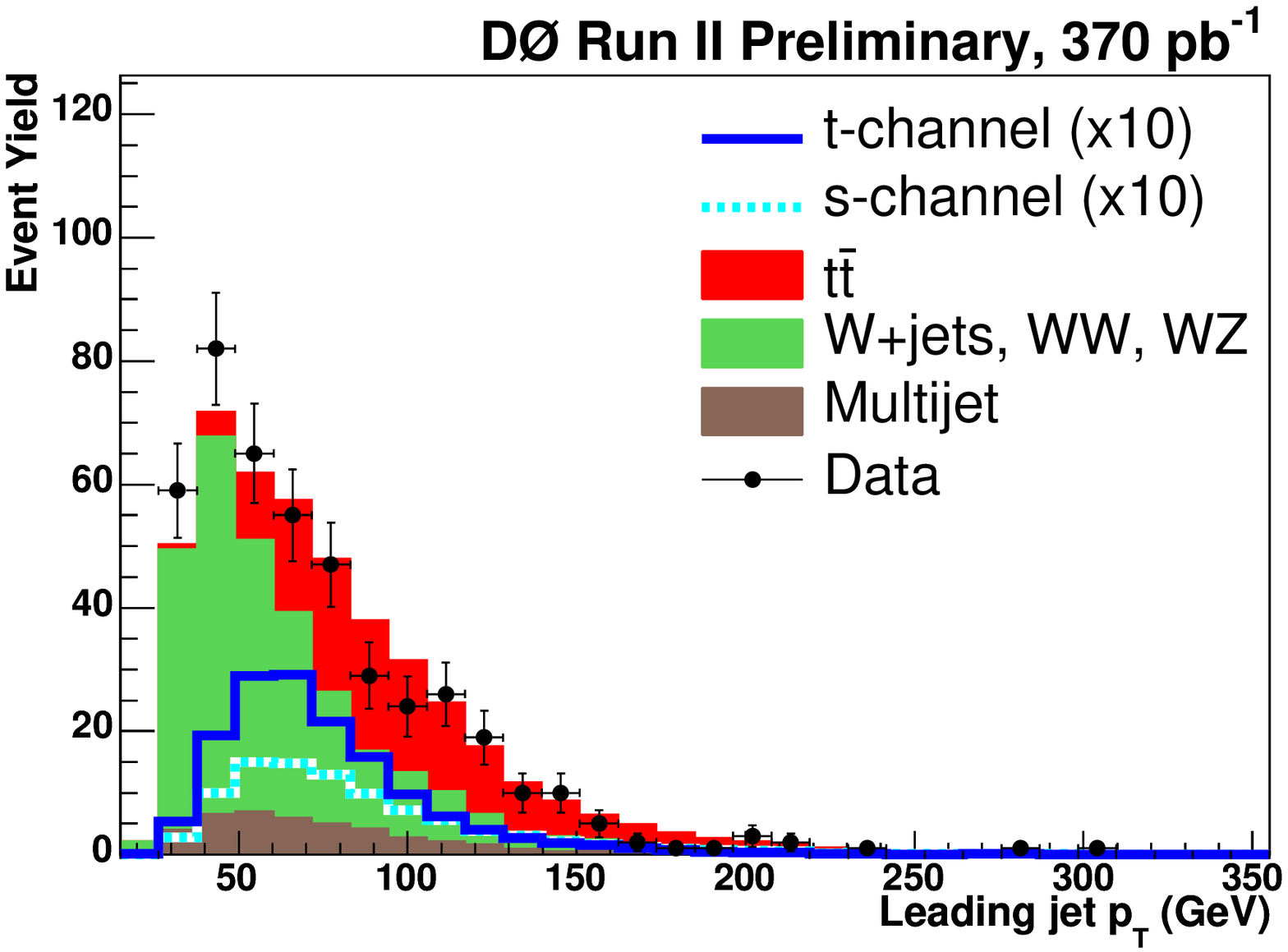}
  \hspace{0.2cm}
  \includegraphics[width=0.36\textwidth]{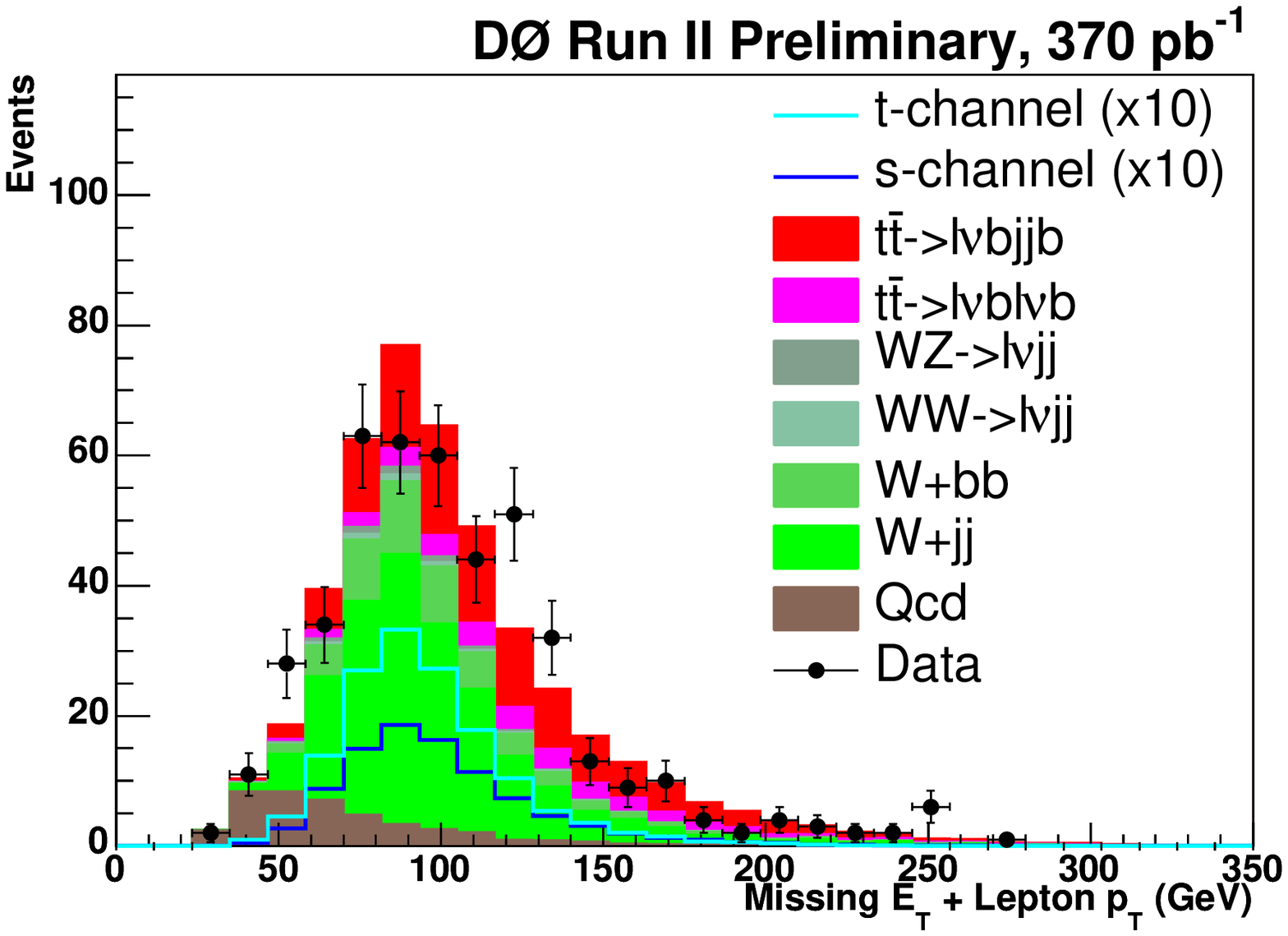}
  \includegraphics[width=0.36\textwidth]{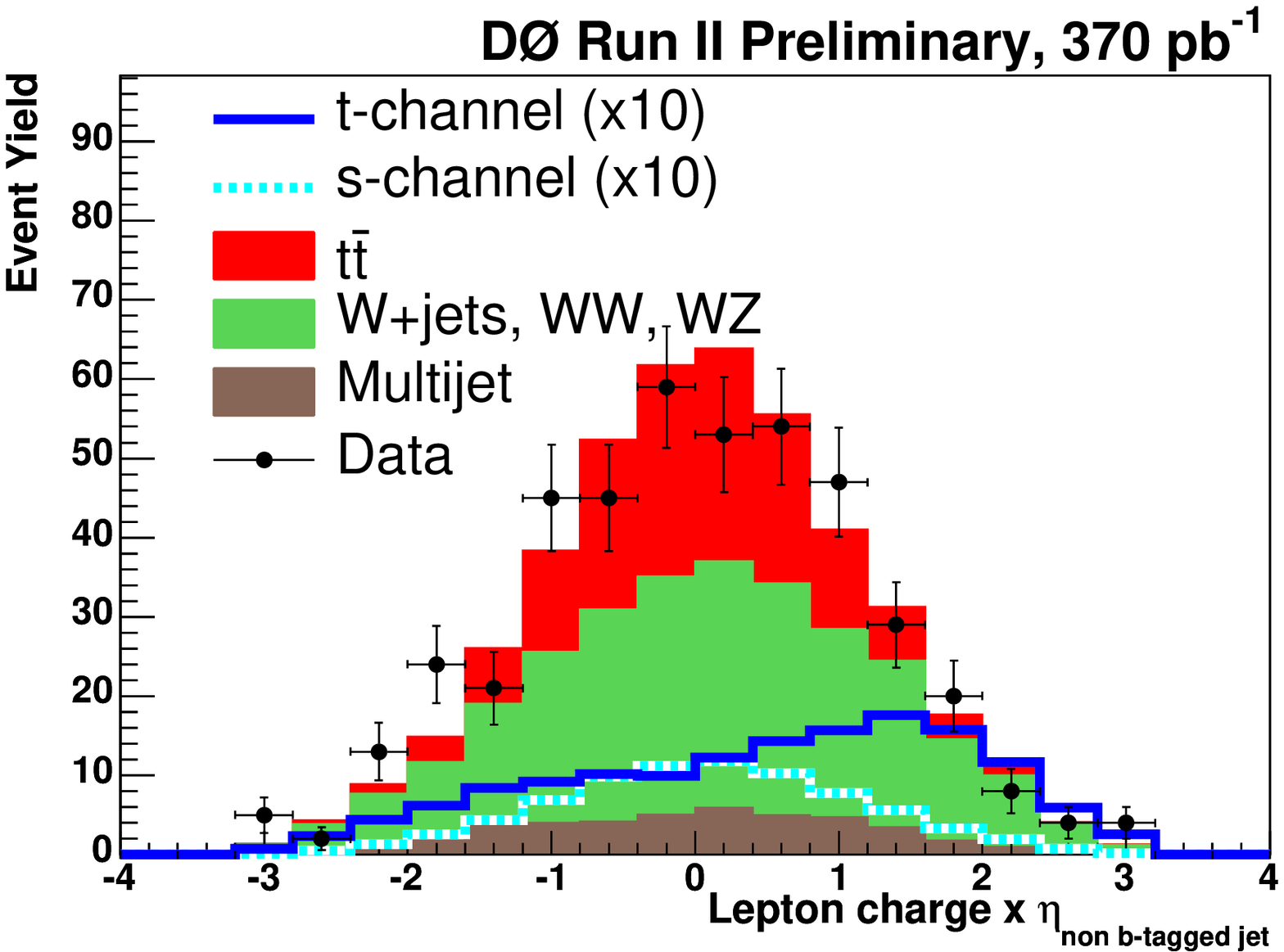}
  \hspace{0.2cm}
  \includegraphics[width=0.36\textwidth]{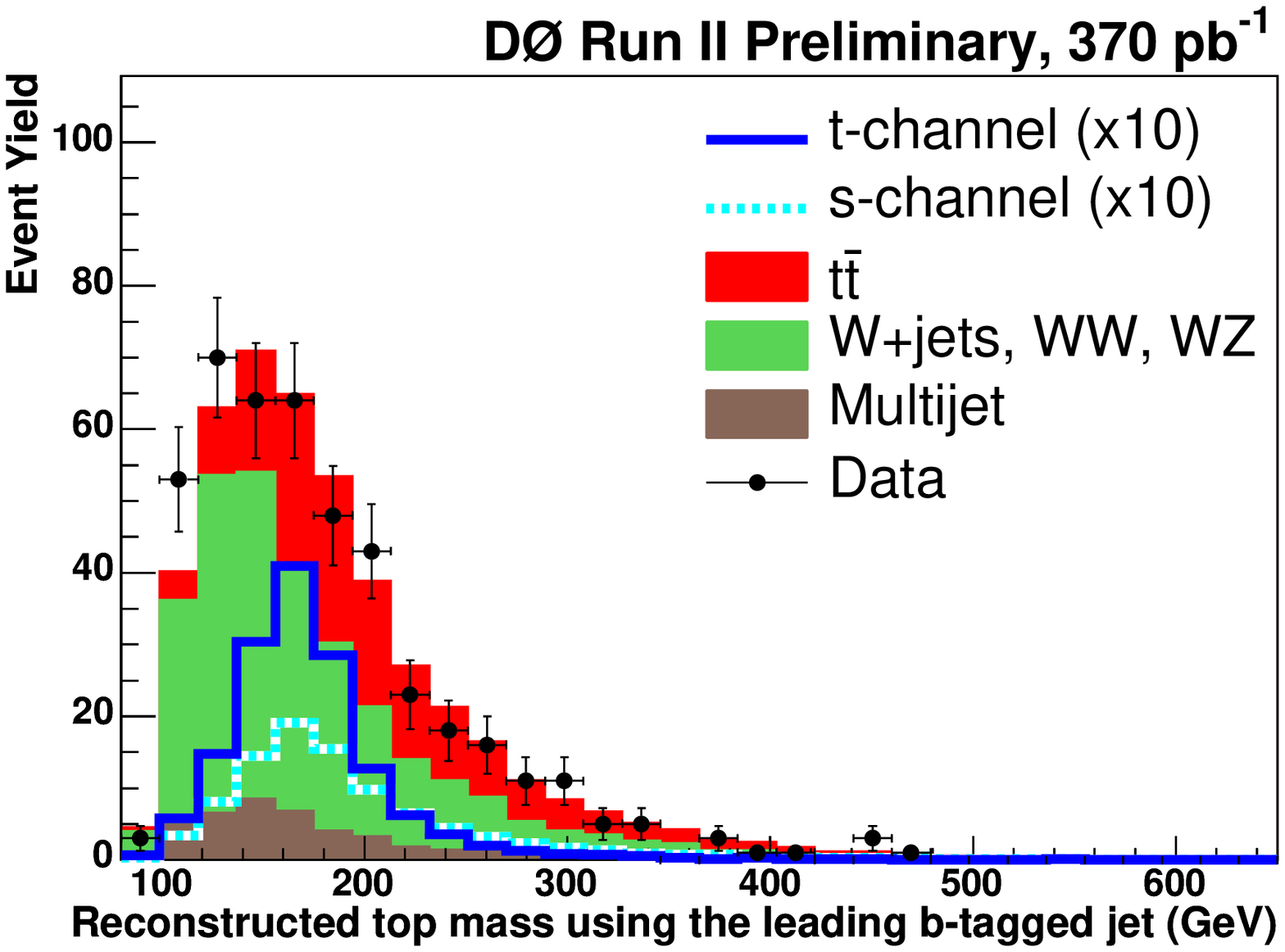}
  \caption{Distributions for data and background with the electron and muon
channels combined of: leading jet $\pt$, the $\met$+lepton $\pt$, the charge of the lepton times the
$eta$ of the untagged jet, and the invariant mass of the reconstructed $W$ and
the leading tagged jet.}
\label{fig:vars}
\end{center}
\end{figure}

At this point in the analysis, cross-check samples enriched in $W$+jets and
$\ttbar$ events only are utilized to convince ourselves that the background
modeling is correct not only in the signal region but also in other regions. 
Figure~\ref{fig:vars} shows some of the variables in the
signal region used later in the analysis. Good agreement is found in all of them
between data and background modeling. 

Single top is a difficult signal to extract from the large backgrounds, as can
be seen from Fig.~\ref{fig:vars}. Since no
single distribution allows to separate the signals from the backgrounds,
multivariate analysis techniques are best suited to efficiently examine a
large portion of the phase space trying to identify and isolate regions were the signal
sticks out. For this analysis we have employed likelihood discriminants with
mostly uncorrelated variables as input. Here is where we benefit from the loose
selection cuts: the likelihood discriminants do a better job of extracting the
signal than any optimization of simple stringent cuts on basic object
distributions. 

\begin{figure}
\begin{center}
  \includegraphics[width=0.36\textwidth]{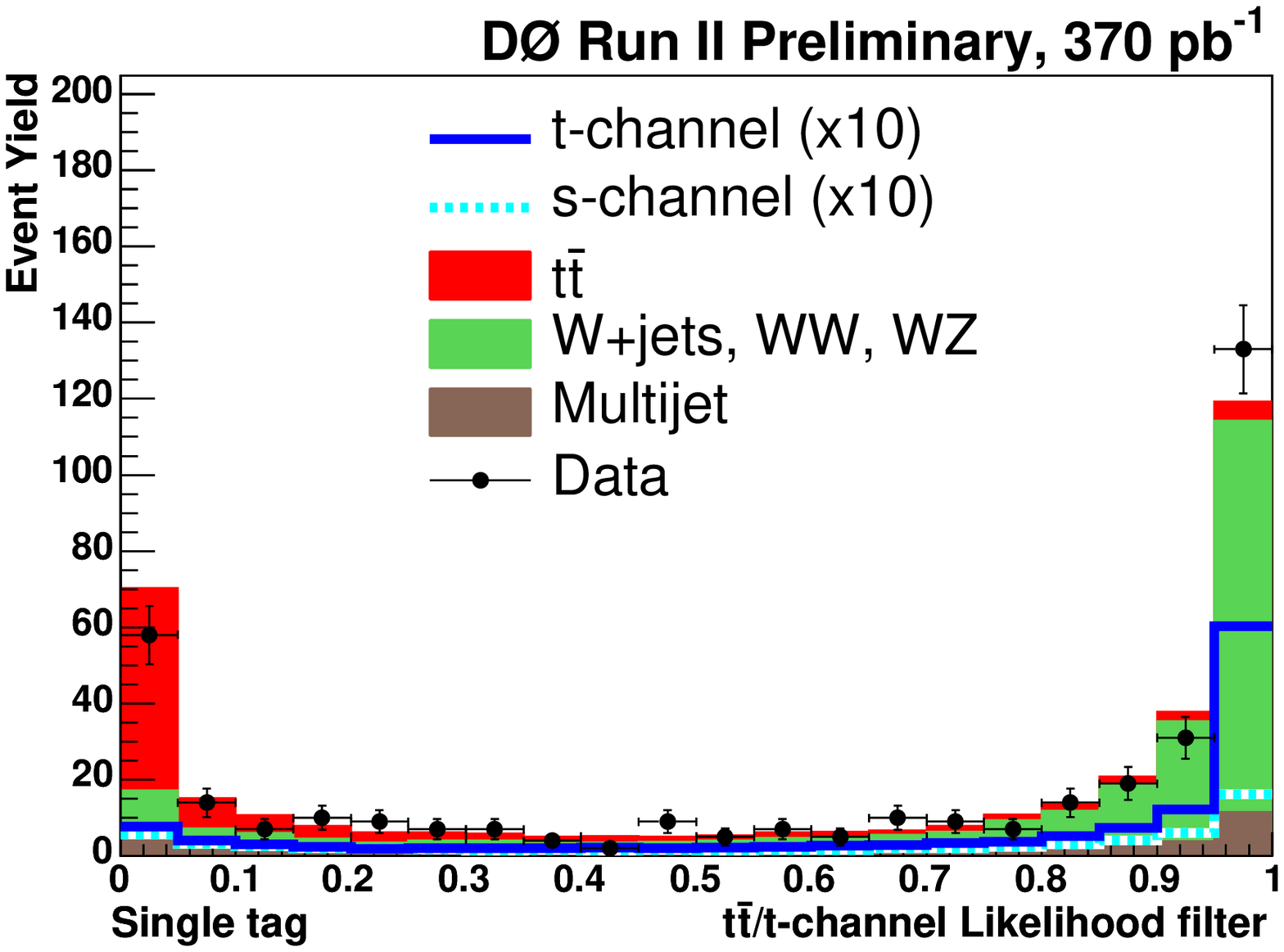}
  \hspace{0.2cm}
  \includegraphics[width=0.36\textwidth]{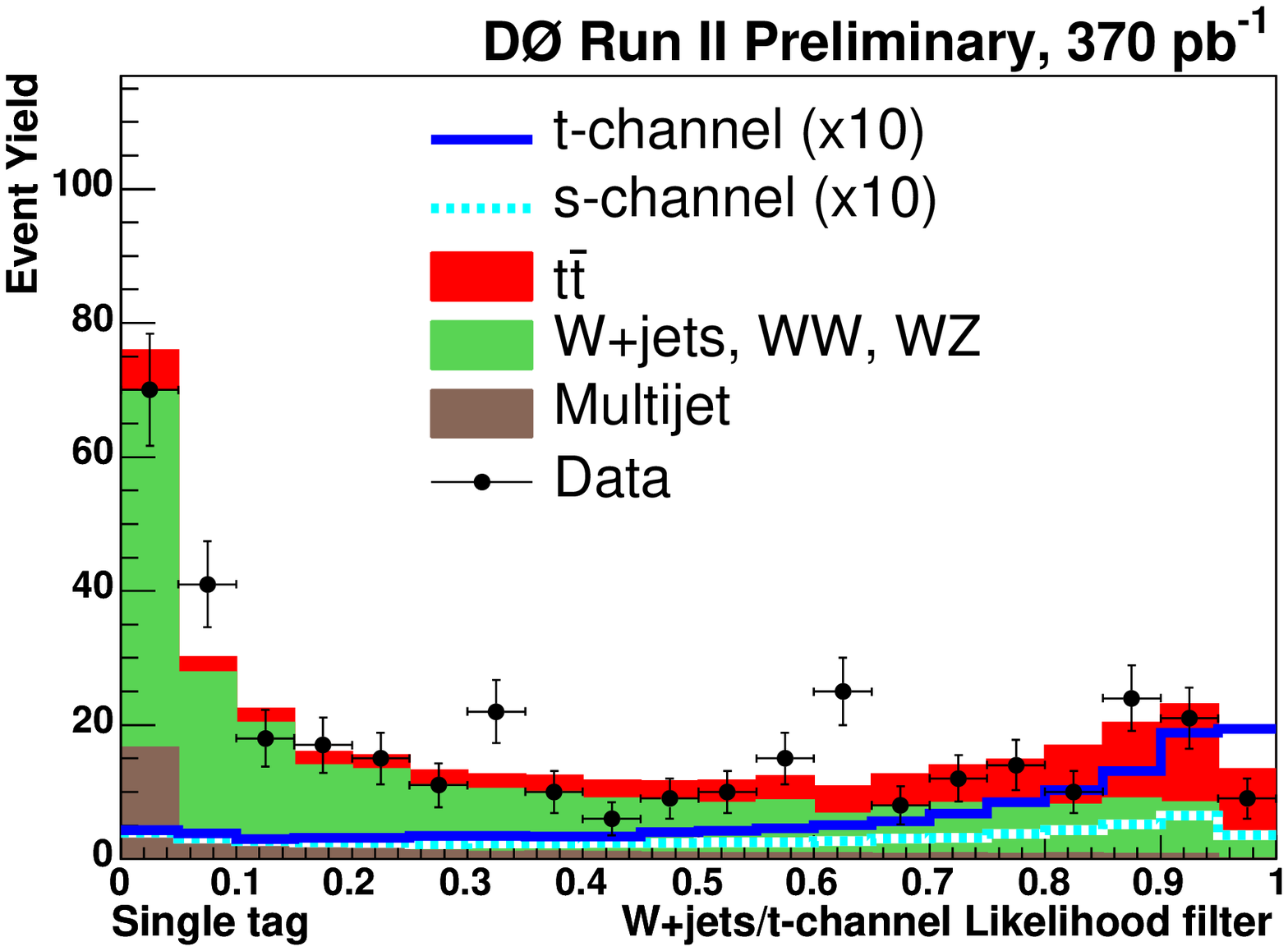}
  \includegraphics[width=0.36\textwidth]{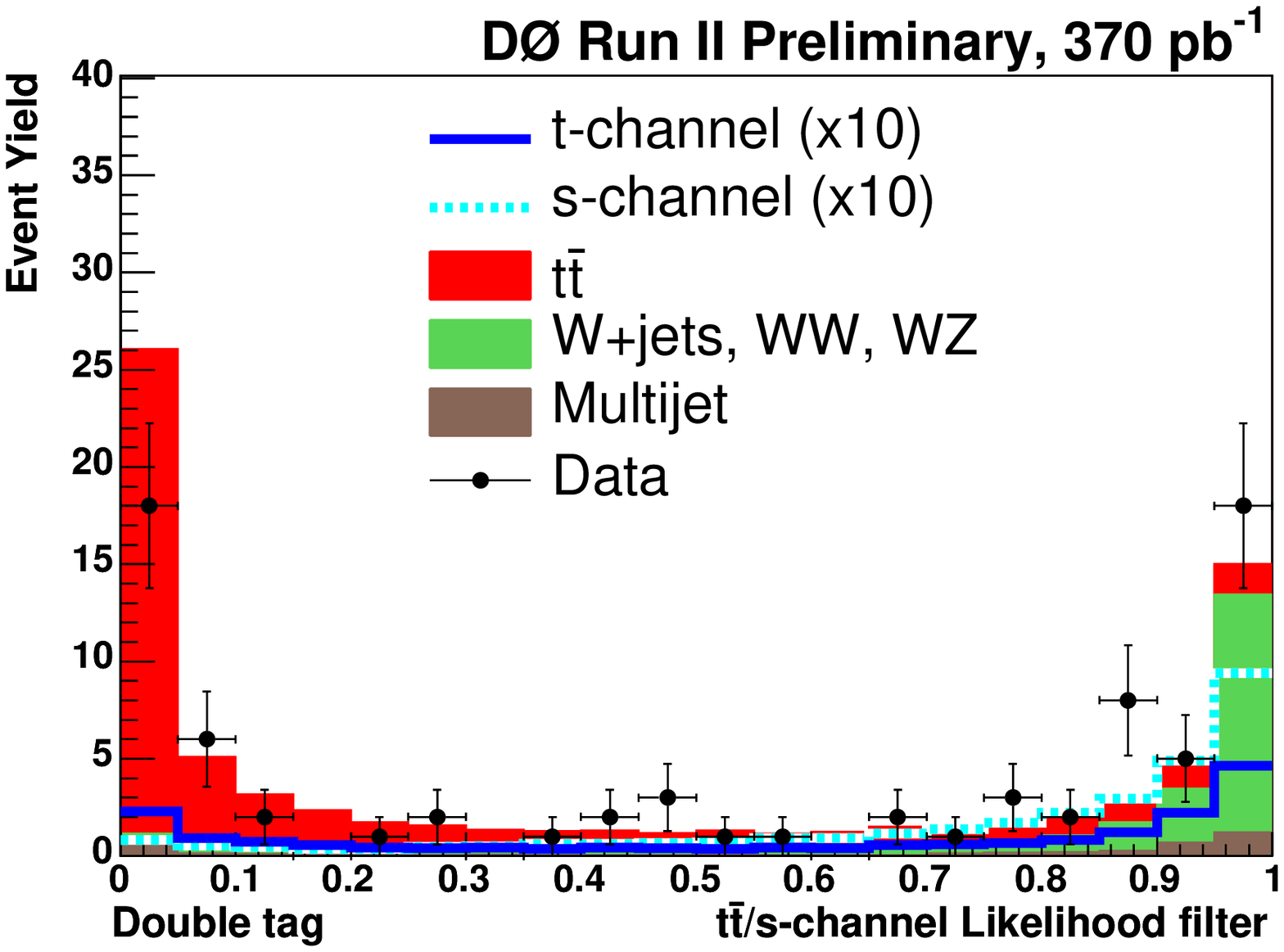}
  \hspace{0.2cm}
  \includegraphics[width=0.36\textwidth]{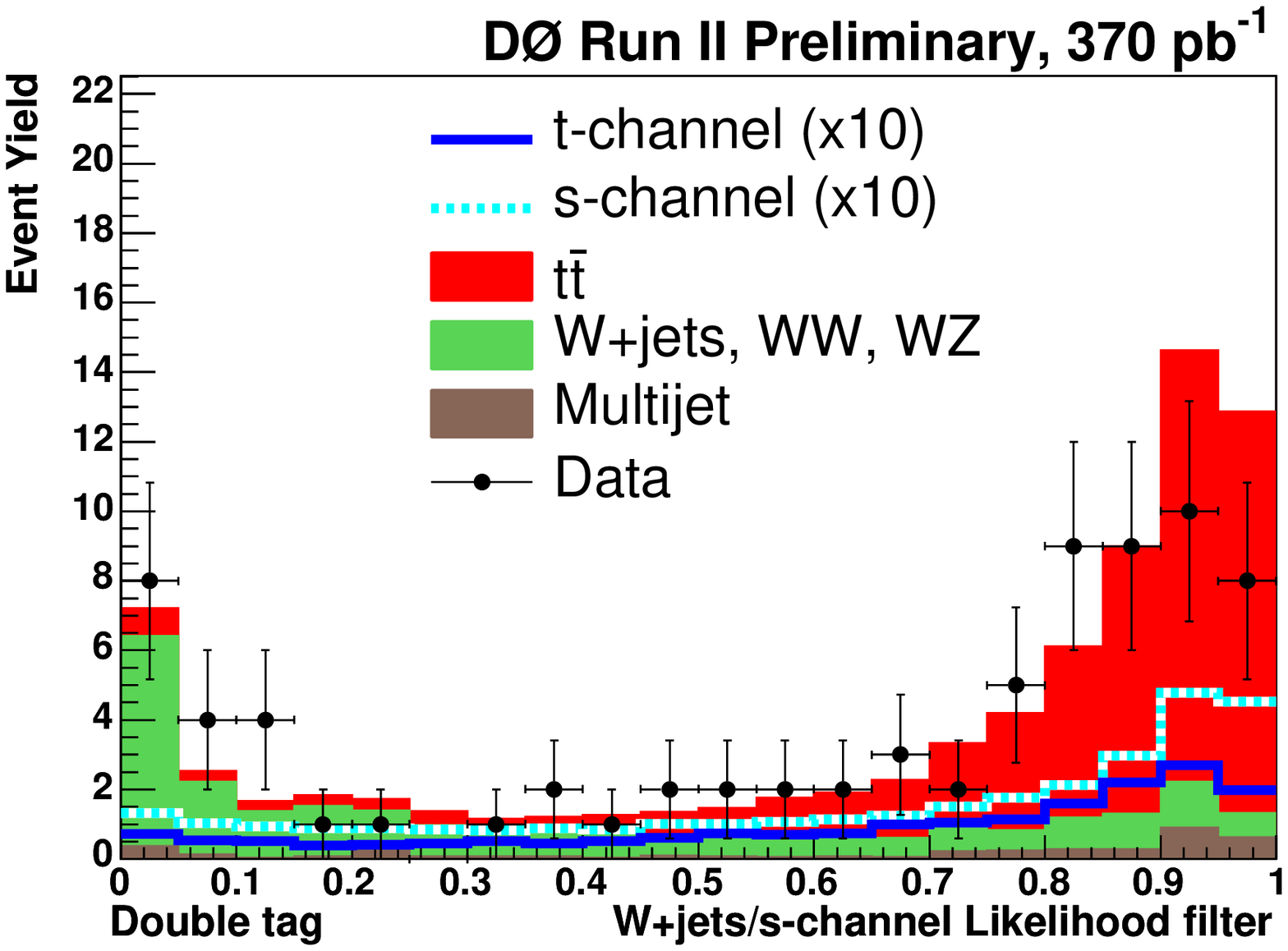}
  \caption{Likelihood outputs for electron and muon channels combined in the
single tagged $tqb$/$\ttbar$ (upper left) and $tqb$/$W$+jets (upper right) filters;
and for double tagged $tb$/$\ttbar$ (lower left) and $tb$/$W$+jets (lower right)
filters.}
\label{fig:lhood}
\end{center}
\end{figure}

The likelihood discriminant is defined as: 
${\cal L}({\vec x}) = \frac{{\cal P}_{\rm signal}({\vec x})}{ {\cal P}_{\rm
signal}({\vec x}) + {\cal P}_{\rm background}({\vec x}) }$,  
where ${\cal P}_{\rm signal}({\vec x})$ and ${\cal P}_{\rm background}({\vec
x})$ are the probability density functions of signal and background events as
derived from the product of input template distributions. We have performed this analysis
in separate channels for electron/muon, single/double tagged, 
$s$-channel/$t$-channel signals and $W$+jets/$\ttbar$ backgrounds giving a total
of 16 different likelihoods. Given the different topologies of the main
backgrounds and signals, we thus benefit from the different separation power
that each likelihood can accomplish in each channel and then can easily combine
them for the extraction of the limit. 
The output of some of the likelihoods can be seen in Fig.~\ref{fig:lhood}.
\section{Results and conclusions}
We use two-dimensional distributions of the output likelihoods of $\ttbar$ $vs.$
$W$+jets for each signal, and use the shape information on each bin of that
plane, taking into account the correlations in the errors in each bin, in a
Bayesian analysis with a Poisson distribution for the observed counts in data and
flat priors for the signal cross sections. The priors for the signal acceptance
and the backgrounds are multivariate gaussians. 
The observed (expected) 95\% confidence level upper limits are 5.0 (3.3)~pb for
the $s$-channel and 4.4 (4.3)~pb for the $t$-channel. 

With more than 1~fb$^{-1}$ of delivered luminosity and better object
reconstruction, calorimeter calibration, much reduced jet energy scale errors
(which improve our invariant mass calculation), better $b$-tagging techniques,
and more powerful separation techniques like matrix element and boosted decision
trees, we can expect the single top search at \dzero~to enter a very exciting
time in the near future. We are at the same time looking for physics beyond the
standard model in this final state topology coming from a new massive $W'$
boson~\cite{wprime} and flavor changing neutral currents~\cite{fcnc}.  

\section*{Acknowledgments}
The author would like to thank the organizers for the hospitable and enjoyable
atmosphere achieved throughout the meeting.


\begin{thebibliography}{99}
\bibitem{d0singletop}
V.M.~Abazov {\it et al}, \Journal{\PLB}{622}{265}{2005}, hep-ex/0505063;
V.M.~Abazov {\it et al}, submitted to Phys. Rev. D. Fermilab-PUB-06/069-E, 
hep-ex/0604020; V.M.~Abazov {\it et al}, \dzero~Public Note 4871.

\bibitem{alpgen}
M.L.~Mangano, M.~Moretti, F.~Piccinini, R.~Pittau, and A.D.~Polosa,
J. High Energy Physics {\bf 0307}, 001 (2003).

\bibitem{comphep}
E.~Boos {\it et al}, \Journal{\NIM}{534}{250}{2004}.

\bibitem{d0nim}
V.M.~Abazov {\it et al}, submitted to NIM A, Fermilab-PUB-05/341-E.

\bibitem{wprime}
V.M.~Abazov {\it et al}, submitted to Phys. Lett. B.
Fermilab-Pub-06/257-E, hep-ex/0607102.

\bibitem{fcnc}
V.M.~Abazov {\it et al}, in preparation. 

\end{thebibliography}
\end{document}